\documentclass[preprint,groupedaddress,amsmath,amssymb,showpacs]{revtex4-1}
\usepackage{graphics}
\usepackage{graphicx}
\usepackage{natbib}
\usepackage{upgreek}
\usepackage{textcomp}

\begin{document}

\title{Tunable microwave impedance matching to a high impedance source using a Josephson metamaterial}

\author{Carles Altimiras}
\email{carles.altimiras@sns.it}
\altaffiliation[Present address: ]{NEST, Istituto Nanoscienze CNR and Scuola Normale Superiore, Piazza San Silvestro 12, 56127, Pisa, Italy}
\affiliation{Service de Physique de l'Etat Condens\'{e} (CNRS URA 2464), IRAMIS, CEA Saclay, 91191 Gif-sur-Yvette, France}
\author{Olivier Parlavecchio}
\affiliation{Service de Physique de l'Etat Condens\'{e} (CNRS URA 2464), IRAMIS, CEA Saclay, 91191 Gif-sur-Yvette, France}
\author{Philippe Joyez}
\affiliation{Service de Physique de l'Etat Condens\'{e} (CNRS URA 2464), IRAMIS, CEA Saclay, 91191 Gif-sur-Yvette, France}
\author{Denis Vion}
\affiliation{Service de Physique de l'Etat Condens\'{e} (CNRS URA 2464), IRAMIS, CEA Saclay, 91191 Gif-sur-Yvette, France}
\author{Patrice Roche}
\affiliation{Service de Physique de l'Etat Condens\'{e} (CNRS URA 2464), IRAMIS, CEA Saclay, 91191 Gif-sur-Yvette, France}
\author{Daniel Esteve}
\affiliation{Service de Physique de l'Etat Condens\'{e} (CNRS URA 2464), IRAMIS, CEA Saclay, 91191 Gif-sur-Yvette, France}
\author{Fabien Portier}
\affiliation{Service de Physique de l'Etat Condens\'{e} (CNRS URA 2464), IRAMIS, CEA Saclay, 91191 Gif-sur-Yvette, France}

\date{\today}


\begin{abstract}
We report the efficient coupling of a $50\,\Omega$ microwave
circuit to a high impedance conductor. We use an impedance transformer
consisting of a $\lambda/4$ co-planar resonator whose inner conductor
contains an array of superconducting quantum interference devices
(SQUIDs), providing the resonator with a large and tunable lineic
inductance $\mathcal{L}\sim 80 \mu_0$, resulting in a large
characteristic impedance $Z_C\sim 1\,\mathrm{k}\Omega$.  
The impedance matching efficiency is characterized by measuring the shot noise power
emitted by a dc biased high resistance tunnel junction connected to the resonator.  
We demonstrate matching to impedances in the $15$ to $35\,\mathrm{k}\Omega$
range with bandwidths above $100\,\mathrm{MHz}$ around a
resonant frequency tunable in the $4$ to $6\,\mathrm{GHz}$ range.
\end{abstract}

\maketitle 

In an electrical circuit, the output impedance of a source must be
matched to the input impedance of the load in order to maximize power
transfer\cite{ArtofElectronics}. In RF circuits, where $50\,\Omega$
matching is standard, it is difficult to detect small signals
emitted from high output impedance sources. For quantum conductors, whose
output impedance is typically on the order of the resistance quantum,
$\mathrm{R_K}=h/e^2\simeq 25.8\,\mathrm{k}\Omega$, matching is
especially difficult.

Developing techniques to solve this mismatch is technologically
important since high output impedance quantum circuits are promising
as ultra sensitive detectors of charge, e.g. Single Electron
Transistors (SETs)
\cite{FultonDolanSETPRL1987,SchoelkopfRFSETScience1998,AassimeRFsetOptimizedAPL2001},
or magnetic flux, e.g. Superconducting Quantum Interference Proximity
Transistors (SquipTs)
\cite{GiazottoSquipTNaturePhys2010,GiazottoSquipTThyPRB842011}. In
both cases, their high output impedance ($\sim100\,\mathrm{k\Omega}$) and a
typical 100 pF stray capacitance to ground of the wiring 
limit the detection bandwidth from dc to a few kHz. An efficient technique used to overcome this limitation, originally
developed to efficiently measure SETs
\cite{SchoelkopfRFSETScience1998}, consists in embedding the sensor in a RF resonant circuit, which allows matching its 
high output impedance to  50 $\Omega$ 
at a chosen frequency. Resonant RF circuits can be made using discrete lumped
elements externally connected to the chip
\cite{SchoelkopfRFSETScience1998,MullertunableRFQPCAPL2010}, or
defined on-chip \cite{XueonChipLCmatchingAPL2007} with transmission
lines terminated by an adequate impedance to ground, e.g. a
stub-shunted
geometry\cite{HellmullerRFvaractMatchAPS2012,Puebla-hellmannStubMatchingAPL2012}. On-chip
impedance matching has the advantage of reducing the stray 
capacitance, allowing to reach detection bandwidths in the 10 MHz range\cite{XueonChipLCmatchingAPL2007,HellmullerRFvaractMatchAPS2012,Puebla-hellmannStubMatchingAPL2012}. 
Here we report on a  technique which offers matching to
high detection impedances (demonstrated up to $35\,\mathrm{k\Omega}$) at a tunable frequency 
in the few GHz range and with bandwidth above $100\,\mathrm{MHz}$. 

As shown in Fig.1(a),  we have developed  $\lambda/4$ co-planar, 
resonators \cite{Pozar} whose inner conductor contains a high kinetic inductance
metamaterial \cite{CastellanosJPAAPL2007}, namely a series array of superconducting quantum
interference devices (SQUIDs). This increases the
characteristic impedance of the resonator\cite{MaslukSuperInductorPRL2012,BellSuperInductorPRL2012} and the resulting detection
impedance. Using SQUIDs instead of plain Josephson junctions allows
tuning in-situ the kinetic inductance by threading a magnetic
flux in the SQUID loops. The resonator matches a highly resistive normal tunnel junction with a tunnel resistance
$R_\mathrm{T}=230\,\mathrm{k}\Omega$ to a 50 $\Omega$ measurement line. We characterize the resonator by detecting the well known shot-noise
produced when a dc voltage bias is applied to the junction.

\paragraph*{•}
A $Z_0= 50 \Omega$ load terminating a section of length $l$ of a transmission line with  characteristic impedance  $Z_1$ and phase velocity $c$ 
is transformed into an impedance depending on the frequency $\nu$\cite{Pozar}:
\begin{equation}\label{Eq1}
Z(\nu)=Z_1 \frac{Z_0+j Z_1 \tan (2 \pi \nu l/c)}{Z_1+j Z_0 \tan (2 \pi \nu l/c)},
\end{equation}

The impedance is real and equal to $Z_1^2/Z_0$ at resonant frequencies $\nu_n=(2n +1) c / 4l$, with $n$ integer, thus transforming the load
impedance $Z_0$ into a higher impedance when $Z_1>Z_0$. For $n=0$, $l$ is equal to a quarter of a wavelength, 
hence the name of quarter wavelength transformer.
  Assuming $Z_1 \gg Z_0$,  the real part of the impedance reads in the vicinity of $\nu_0$:
 $\mathrm{Re}[Z(\nu)] \sim \displaystyle{\frac{Z_1^2}{Z_0}}
\displaystyle{\Big(1+\Big(\frac{\pi Z_1 (\nu-\nu_0)}{2 Z_0\nu_0}\Big)^2\Big)}^{-1}.$
It is equivalent to a damped LC circuit with a resonant frequency
$\nu_0=1/(2 \pi \sqrt{\mathrm{LC}})$,  impedance
$Z_\mathrm{\mathrm{Res}}=\sqrt{\mathrm{L}/\mathrm{C}}=4Z_1/\pi$. The external quality
factor, due to dissipation in the matched
load $Z_0$, is  $Q=\frac{\pi}{4}\frac{Z_1}{Z_0}$. Reaching an
impedance $Z(\nu_0)\simeq R_\mathrm{K}$ requires a transmission line with
$Z_1\simeq 1\,\mathrm{k}\Omega$. This is well beyond the values attainable with metallic co-planar transmission lines, for which $Z_1$ only depends logarithmically on 
their transverse dimensions, but is achievable with high kinetic inductance lines.

 A balanced SQUID
\cite{JaklevicSQUIDPRL1964} consists in two identical parallel Josephson
junctions closing a superconducting loop. When the inductance of the loop connecting
the Josephson junctions is negligible, as here, its inductance is $L_\mathrm{J}(\phi)=\phi_0/I_C(\phi)$
with  $\phi_0= \hbar/2e$ , $\phi$ is the flux applied through the SQUID, and $I_C$ the critical current of the SQUID. $I_C$ is given by the Ambegaokar-Baratoff formula \cite{AmbegaokarBaratoffPRL1963}, 
 $I_C=\frac{\pi\Delta|\cos(\phi/2\phi_0)|}{e R_\mathrm{N}}$  with  $R_\mathrm{N}$ is the
tunnel resistance of each junction. This expression for the Josephson inductance is valid provided that 
i)  the characteristic energies $h \nu$ and $k_\mathrm{B} T$ respectively associated to the frequency and the temperature are much smaller than the superconducting gap,
 ii) the phase across the junction can be considered as a classical quantity\cite{Joyez2013}, and iii) is much smaller than 1, which implies that the current going through the SQUID is much smaller than $I_C$. 
To evaluate the lineic impedance $\mathcal{Z}$ of the inner conductor of our transmission line, 
one must take into account the capacitance $C_\textrm{J}$ of
the tunnel junction, yielding  $\mathcal{Z}= i 2\pi \nu L_\mathrm{J}(\phi) /a (1-\nu^2/\nu_\textrm{P}^2)$, with
$\nu_\textrm{P}$ the Josephson plasma frequency\cite{JosephsonRMP1964}
$\nu_\textrm{P}=1/[2\pi(L_\textrm{J}C_\textrm{J})^{1/2}]$, and $a$ is the distance between neighbouring SQUIDs. 
Below $\nu_\textrm{P}$, the inner conductor thus presents an effective lineic inductance 
\begin{equation}
\mathcal{L}= L_\mathrm{J}(\phi) /a (1-\nu^2/\nu_\textrm{P}^2).
\label{Jeff}
\end{equation}
The characteristic impedance of the SQUID transmission line is then given by  $Z_1=\sqrt{\mathcal{L}/\mathcal{C}}$, where $\mathcal{C}$ 
is the lineic capacitance to ground of the SQUID array, evaluated using standard electromagnetic simulations. Note that at frequencies much lower than  
$\nu_\textrm{P}$, $\mathcal{L}$ is independent of frequency, so that our device can be considered as a standard quarter wavelength impedance transformer, albeit with
high lineic inductance. Finally, we would like to point that this technique cannot provide 
 characteristic impedances $Z_1 \gg R_\textrm{Q}$, since quantum phase slips then drive the array into an insulating state
\cite{ChowSuperToInsulArraysPRL1998}.

Figure 1(a) shows a picture of the measured co-planar resonator and its
equivalent electric circuit scheme. From the left to the right a
$50\,\Omega$ line is followed by a $350\,\mathrm{\upmu m}$ long
Josephson metamaterial line containing lithographically identical and
evenly spaced SQUIDs with a $5\,\mathrm{\upmu m}$ period. SQUIDs
where fabricated following the process described in Ref.
\cite{PopReproducibleJunctionsArXiv2012}: the SQUIDs
(see the top inset) are obtained by double angle deposition of
($20/40\,\mathrm{nm}$) thin aluminum electrodes, with a $20'$
oxydation of the first electrode at $400\,\mathrm{mBar}$ of a ($85\%\,\mathrm{O}_2/15\%\,\mathrm{Ar}$) mixture. 
Before the evaporation, the substrate was cleaned by rinsing in ethanol and Reactive Ion Etching in an oxygen plasma \cite{Plasma}.  
The barriers have an area of
$0.5\,\mathrm{\upmu m^2}$ each resulting in a room temperature
tunnel resistance $R_\mathrm{N}=720\,\Omega$. To assess that the SQUIDs in the array are identical, we have perfomed reproducibility tests,
yielding constant values of $R_\mathrm{N}$ (within  a few $\%$) over millimetric distances. Assuming a superconducting gap $\Delta=180\,\mathrm{\upmu
  eV}$ and a 17\% increase of the tunnel resistance between room temperature and base temperature\cite{GloosGTTunnelAPL}, one obtains a zero flux critical current for the SQUIDs $I_C= 671\,\mathrm{nA}$, 
  corresponding to $L_J(\phi=0)=0.49\,\mathrm{nH}$. 
 This corresponds to an effective lineic inductance   $\mathcal{L}\simeq\mathrm{100\,\upmu H.m^{-1}}$ at zero
magnetic flux and frequency much lower than $\nu_\textrm{P}$. Assuming  a capacitance for the junctions of the order of $80\,\mathrm{fF}/\upmu \mathrm{m}^2$ yields $\nu_\textrm{P}\simeq 25\,\mathrm{GHz}$.
Note that our simple fabrication mask (see Top Panel in Fig. 1(a)) produces 10 times bigger
Josephson junction in between adjacent SQUIDs, resulting in an additional $\sim \mathcal{L}\simeq\mathrm{10\,\upmu H.m^{-1}}$ lineic  inductance.
The  $\sim \mathcal {L}\simeq\mathrm{1\,\upmu H.m^{-1}}$ electromagnetic inductance associated to our geometry is negligible. 
  With the designed lineic capacitance
$\mathcal{C}=84.3\,\mathrm{pF.m^{-1}}$, the length of the resonator
sets the first resonance at $\nu_0\simeq7\,\mathrm{GHz}$ with an
impedance $Z_\mathrm{Res}\simeq 1.5\,\mathrm{k\Omega}$ and quality
factor $Q\simeq 18$. The frequency $\nu_0$ being small enough compared to $\nu_\textrm{P}$,
 the frequency dependence of $\mathcal{L}$ given by Eq.\ref{Jeff} is almost negligible. Sweeping the field then modifies
$\mathcal{L}\propto |\cos (\phi/2\phi_0)|^{-1}$, $\nu_0\propto |\cos (\phi/2\phi_0)|^{0.5}$, the impedance $Z_\mathrm{Res} \propto |\cos
(\phi/2\phi_0)|^{-0.5}$ and the quality factor $Q \propto |\cos
(\phi/2\phi_0)|^{-0.5}$.  The $0.35\,\mathrm{mT}$ field period is estimated from the 
$3\,\mathrm{\upmu m^2}$ SQUID area. 

The  $100\times
100\,\mathrm{nm^2}$ normal tunnel junction (see Fig. 1(a) bottom inset) is fabricated by multiple angle
evaporation of copper/aluminum/copper with thicknesses ($30/5/60\,\mathrm{nm}$), and
a $20'$ oxydation of aluminum at $800\,\mathrm{mBar}$ of a
($85\%\,$O$_2/15\%\,$Ar) mixture. Using the standard theory of the proximity effect \cite{DeGennesRevModPhys1964}, we find that the
aluminum  superconductivity is fully suppressed in the lower electrode.
Finally  a $30 \times 50 \times 0.3\, \mathrm{\upmu m^3}$ normal metal (gold) pad is inserted between the SQUID array 
and the normal junction, in order to efficiently absorb the  power dissipated at the biased normal tunnel junction. The electron-phonon
coupling \cite{HuardHeatingFilsmPRB2007,GiazottoRevModPhys2006} is sufficient to maintain the  
 electronic temperature below $\sim 20\,\mathrm{mK}$ at the maximum bias voltage used here. For
the same reason, the ground plane is also made of
$0.3\,\mathrm{\upmu m}$ thick gold. This pad adds an extra capacitance to ground $C_{\mathrm{shunt}}\simeq
12\,\mathrm{fF}$ (estimated numerically).  The impedance across the tunnel junction $Z(\nu)$ is the parallel combination of $C_{\mathrm{shunt}}$ and  the 
50$\ \Omega$ load transformed by the quarter wavelength section. 
As seen from the junction, the main effect of
this capacitance is thus to reduce the resonant frequency and the associated impedance ($\nu_0\simeq 6\,\mathrm{GHz}$ and $Z_\mathrm{\mathrm{Res}} \simeq 1 \ \mathrm{k}\Omega$ at zero flux).

A schematic diagram of the measuring set-up is shown in  Fig. 1(b). The device is cooled in a dilution fridge with a $T=15\,\mathrm{mK}$ base temperature.
The device is connected to the measurement circuit through a $50\,\Omega$ matched commercial bias Tee. 
Its inductive port is used to current bias the normal tunnel junction via a $13\,\mathrm{M}\Omega$ 
resistor anchored at $0.8\,\mathrm{K}$. The resulting dc voltage across the sample is measured in a 3 point configuration via a room 
temperature commercial voltage pre-amplifier, which allows an in-situ measurement of the voltage across the device, from which we deduce $R_\mathrm{T}$. The dc lines are carefully filtered by custom made passive filters (a $3^{\mathrm{rd}}$ order RC filter with 
$R=2\,\mathrm{k}\Omega$, $C=1\,\mathrm{nF}$ anchored at $0.8\,\mathrm{K}$, followed by a $170\,\Omega$, 
$450\,\mathrm{pF}$ distributed RC filter and a bronze powder filter, both anchored at base temperature). 

The capacitive port of the bias Tee couples the sample  to  two RF lines   through 
a $20\,\mathrm{dB}$ directional coupler, whose direct port  couples the sample to the 
detection chain. This chain  consists of a $4-8\,\mathrm{GHz}$ isolator,  a $4-8\,\mathrm{GHz}$ band pass filter,  
a  low pass dissipative gaussian filter ($-3\,\mathrm{dB}$ at $12\,\mathrm{GHz}$) , and   a $40\,\mathrm{dB}$ gain cryogenic  amplifier  with 
$3.5\,\mathrm{K}$ noise  temperature. The rejection noise of the amplifier 
is routed  by the isolator to a $50\,\Omega$ load anchored at the base temperature. The amplified signal is 
then measured at room temperature via an active circuit equivalent to  $15\,\mathrm{MHz}$ band-pass filter 
around   an adjustable frequency $\nu$. This is performed  by  heterodyning the signal  with an external local oscillator using a mixer, and 
passing  the down-converted signal through a low-pass filter. The signal is then fed to a $50 \,\Omega$ 
matched square-law detector whose output dc voltage is proportional to the incoming RF power. Since the resulting signal contains a small 
contribution from the sample on top of the much larger noise  of the cryogenic amplifier,   
a standard Lock-in technique is used  by adding a small ac  bias to the normal tunnel junction with a near dc frequency ($\sim15\,\mathrm{Hz}$) 
and measuring the response of the square law detector in phase with the excitation. 
The gain of the detection chain $G(\nu)$ is calibrated in-situ by exploiting the feed line coupled to the sample via the $-20\,\mathrm{dB}$
port of the directional coupler. This feed line contains $70\,\mathrm{dB}$ attenuation distributed in 3 stages thermally anchored at progressively lower
temperatures, a low pass dissipative gaussian filter ($-3\,\mathrm{dB}$ at $12\,\mathrm{GHz}$), and a $4-8\,\mathrm{GHz}$ bandpass filter,  
which, altogether with the directional coupler, have been independently calibrated at $4.2\,\mathrm{K}$. 
The feed this line is connected to a synthesizer with a level $\sim 1\,\text{dBm}$.

By setting  the device resonance frequency below $4\,\mathrm{GHz}$,  the entire power incoming from the feed line is reflected by the device,
which allows to calibrate the detection chain in the 4-$7 \,\mathrm{GHz}$ frequency range used in the experiments described below. One has to correct for the double passage through 
the bias Tee and the $10\,\mathrm{cm}$ coaxial cable connecting it to the sample, both independently 
calibrated.
With this calibration in hand, we characterize the resonator with the
on-chip radiation source provided by the electronic shot-noise of the tunnel junction. 
Indeed, at large dc bias voltage $eV \gg k_\mathrm{B}T, h\nu$,
a biased tunnel junction is a source of white current noise with a
power spectral density 
$S_\mathrm{I}=2eV/R_\mathrm{T}$. The power density delivered to the $50 \ \Omega$ chain
reads
\begin{eqnarray}
P(\nu)=\frac{2eV\mathrm{Re}[Z(\nu)]R_\mathrm{T}}{|Z(\nu)+R_\mathrm{T}|^2} \simeq \frac{2eV\mathrm{Re}[Z(\nu)]R_\mathrm{T}}{(\mathrm{Re}[Z(\nu)]+R_\mathrm{T})^2},
\end{eqnarray}
 where the last approximation holds if
 $\mathrm{Im}[Z(\nu)] \ll R_\mathrm{T}$. This approximation, which allows to extract $\mathrm{Re}[Z(\nu)]$ from $P(\nu)$,  results in a less than $2\%$ error,
 smaller than the  $\sim
 1-2\,\mathrm{dB}$ uncertainty in the gain calibration. 
 
The impedance $\text{Re}[Z(\nu, \phi)]$
extracted in this manner is shown in the right panel of Fig.2, next to the values predicted without fitting parameter.
Besides a spurious resonance at $5.8\,\mathrm{GHz}$, the data  mostly
present the expected resonance  at  frequency $\nu_0|\cos (\phi/2 \phi_0)|^{0.5}$ with
$\nu_0=6\,\mathrm{GHz}$, and reach detection impedances always above $10\,\mathrm{k}\Omega$.

In order to analyze more quantitatively the impedance matching
capability of our device and its comparison to the model, 
several curves obtained for distinct external magnetic fields are shown in Fig.3
 and compared to the  predictions. These data 
  show a variation of the bandwidth in the    $100$ to $240\,\text{MHz}$ range,
and of the detection impedance  in the  $15$ to $35\,\text{k}\Omega$ range, as predicted. The
  $15\%$ agreement obtained  for the impedance   is compatible with our gain
calibration uncertainty $1-2\,\text{dB}$. 

Beyond an increased coupling to high output impedance sensors, the resonators developed 
here  enrich the toolbox of quantum electrical engineering,  in particular for circuit cavity quantum electrodynamics. Indeed, higher characteristic impedances  
would increase the coupling of double quantum dot qubits to the microwave cavity in which  they are embedded \cite{DelbecqPhysRevLett107-256804-2011,FreyPhysRevLett108-046807-2012},
helping to reach the strong coupling regime. They could also be used to probe Kondo physics with microwave photons \cite{LeHurPhysRevB85-140506-2012,Goldstein-PhysRevLett110-017002-2013}.

Being based on well known technology, it is fairly simple to predict the response of our devices as long as
 their  Josephson junctions remain in the linear regime, which implies that the current must be kept well below their critical current. Note that 
 when approaching the Josephson plasma frequency ($\sim 25\,\mathrm{GHz}$ for aluminum  based junctions), this strongly reduces the 
corresponding maximum microwave power. Above the Josephson plasma frequency, the capacitance of the junctions shunts their kinetic inductance,
 and the basic mechanism of our tecnhique breaks down. This physical limit can however be easily pushed up by using higher gap material, e.g. niobium nitride for which plasma frequencies above a few $100\,\mathrm{GHz}$ have been demonstrated\cite{Villegier, Nagai}.  Other schemes could also be implemented to bypass this limit,
such as using thin films of highly disordered superconductors \cite{DriessenPRL109-107003-2012} which have a 
large  kinetic inductance, or  superconductor-normal
metal-superconductor weak links \cite{JClarkePhysRevB4-2963-1971} which furthermore have a negligible inter-electrode
capacitance.  

In conclusion, we have demonstrated impedance matching to $50\,\Omega$
 of  a high impedance component using a   
microwave resonator embedding a Josephson metamaterial that 
yields a large   resonator  impedance.  The matching frequency is tunable in the   $4 \text{ and } 6\,\mathrm{GHz}$  range,
with  the bandwidth  varying from    $\sim 100$ to $240\,\mathrm{MHz}$.   This impedance matching has been  characterized
using an   on-chip white noise source. This method offers interesting perspectives for quantum electrical engineering at even larger frequency and impedance values.

\paragraph*{}
This project was funded by the CNano-IDF Shot-E-Phot, the Triangle de la Physique DyCoBloS and ANR AnPhoTeQ and Masquel grants. Technical assistance from Patrice Jacques, Pierre-François Orfila and Pascal S\'enat are gratefully acknowledged.

\begin{figure}[floatfix]
  \includegraphics{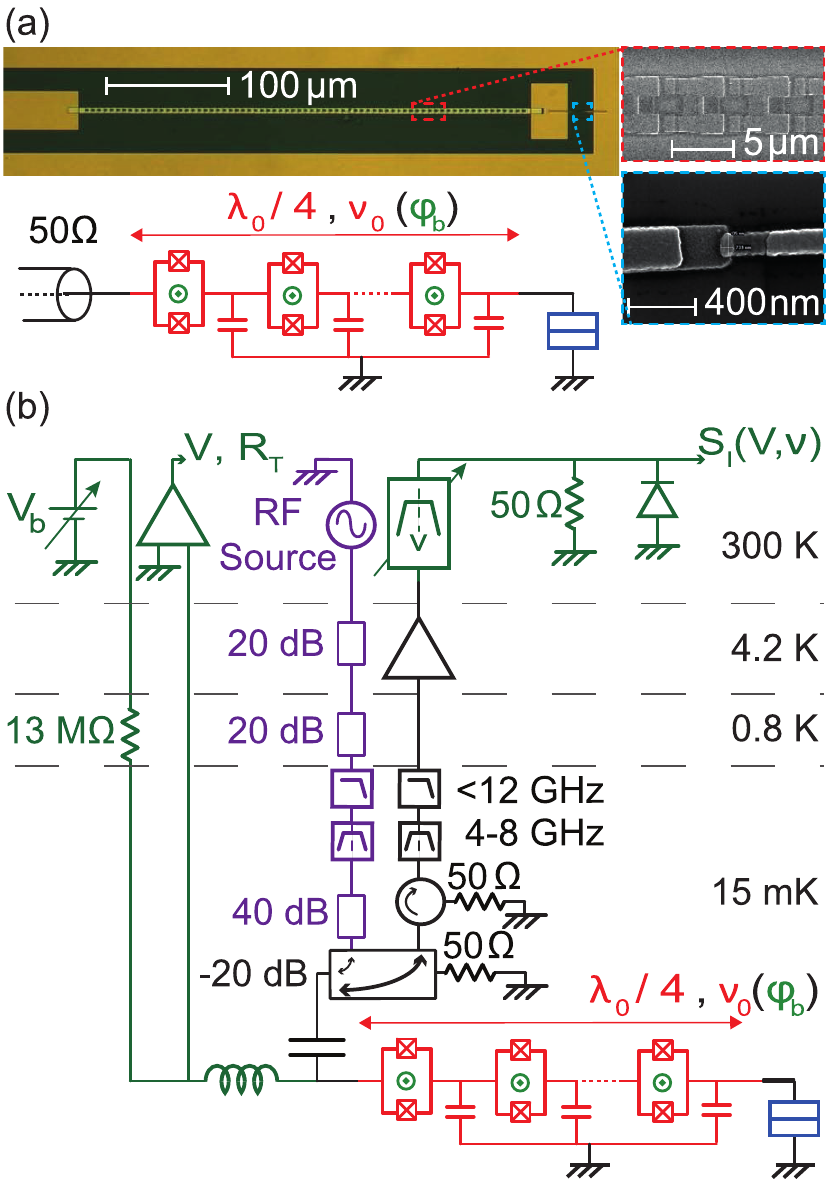}
  \caption{\label{Figure1} (Color online) \textbf{Description of the device and characterization circuit}. \textbf{(a)} Optical microscope image of the $\lambda/4$ co-planar resonator based on a Josephson metamaterial, on top of the scheme of its equivalent electric circuit. Top inset: scanning electron microscope (SEM) image of few SQUIDs from the Josephson metamaterial. Bottom inset: SEM image of the high resistance normal tunnel junction terminating the resonator. \textbf{(b)} Electric circuit used to characterize the resonator: The sample shown in (a) is connected to the circuit with a bias tee. The inductive port is used to dc bias the sample and to measure the normal state resistance of the tunnel junction in a three point configuration. The capacitive port is used to amplify, filter and measure the fraction of the RF power emitted by the tunnel junction which is transmitted by the resonator. This quantity depends only on the dc bias, the detection impedance provided by the resonator, the output impedance of the tunnel junction and the gain of the detection chain. An additional RF line, heavily attenuated and connected to the circuit via a -20dB directional coupler, is used to calibrate the gain of the detection chain. For this calibration, the resonant frequency of the resonator is tuned out of the detection bandwidth so that the incoming radiation is fully reflected by the sample.}
\end{figure}

\begin{figure}[floatfix]
  \includegraphics{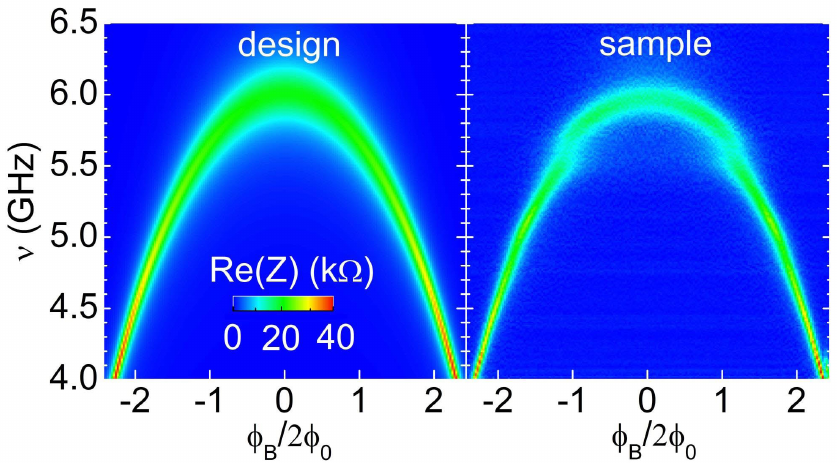}
  \caption{\label{Figure2} (Color online) Left pannel: predicted real part of the detection impedance $\text{Re}[Z(\nu, \phi)]$ obtained by terminating the resonator with a $50\,\Omega$ load as a function of frequency $\nu$ and the externally applied magnetic flux $\phi$. Right pannel: $\text{Re}[Z(\nu, \phi)]$ obtained experimentally. The dc biased tunnel junction is used as an on-chip calibrated RF source feeding the resonator with white current noise. The fraction of RF power transmitted by the resonator depends only on the dc bias, the detection impedance provided by the resonator, the output impedance of the tunnel junction and the gain of the detection chain allowing to extract $\text{Re}[Z(\nu, \phi)]$ with Eq. (3).}
\end{figure}

\begin{figure}[floatfix]
  \includegraphics{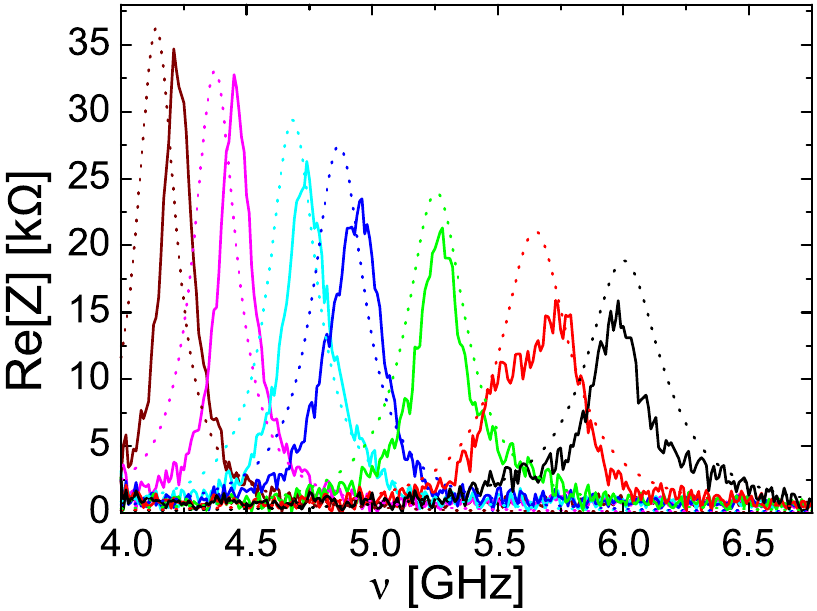}
  \caption{\label{Figure3} (Color online) Comparison between the measured (full lines) and the predicted (dotted lines) $\text{Re}[Z_\text{det}(\nu, \phi)]$ for several values of the externally applied magnetic flux $\phi=(0, 0.16, 0.23, 0.28, 0.3, 0.33, 0.35)$ from the left to the right.}
\end{figure}


\begin{thebibliography}{}
\bibitem [{Pla()}]{Plasma}%
  \BibitemOpen
  \href@noop {} {}\bibinfo {note} {Following $\mathrm{Ref.}^{20}$, we performed
  Reactive Ion Etching under an oxygen pressure of $30\,\mathrm{mBar}$ and
  $10\,\mathrm{W}$ RF power, during 15 seconds, in a Plassys MG -200- S RIE
  equipment.}\BibitemShut {Stop}%
\end{thebibliography}
\end{document}